\documentclass[aps,pra,twocolumn,superscriptaddress]{revtex4}
\usepackage{bm}
\usepackage{graphicx}
\usepackage{amssymb,amsmath,amsbsy,amsgen,amsfonts}
\usepackage{dcolumn}
\usepackage{amsthm}
\usepackage{mathrsfs}
\usepackage{latexsym}
\usepackage{array}
\usepackage{amstext}
\usepackage{epsfig}
\usepackage{epstopdf} 

\usepackage{color}
\usepackage{units}
\usepackage{calrsfs}

\newcommand{\be}{\begin{equation}}
\newcommand{\ee}{\end{equation}}
\newcommand{\ba}{\begin{array}}
\newcommand{\ea}{\end{array}}
\newcommand{\bqa}{\begin{eqnarray}}
\newcommand{\eqa}{\end{eqnarray}}

\begin{document}

\title{Violations of the Bulk-Edge Correspondence in Topological Photonics}
\title{Physical Violations of the Bulk-Edge Correspondence in Topological Electromagnetics}

\author{S. Ali Hassani Gangaraj} \email{ali.gangaraj@gmail.com}
\address{School of Electrical and Computer Engineering, Cornell University, Ithaca, NY 14853, USA}

\author{Francesco Monticone} \email{francesco.monticone@cornell.edu}
\address{School of Electrical and Computer Engineering, Cornell University, Ithaca, NY 14853, USA}

\date{\today}

\begin{abstract}

In this Letter, we discuss two general classes of apparent violations of the bulk-edge correspondence principle for uniform topological photonic materials, associated with the asymptotic behavior of the surface modes for diverging wavenumbers. Considering a nonreciprocal plasma as a model system, we show that the inclusion of spatial dispersion (e.g., hydrodynamic nonlocality) formally restores the bulk-edge correspondence by avoiding an unphysical response at large wavenumbers. Most importantly, however, our findings show that, for the considered cases, the correspondence principle is \emph{physically} violated for all practical purposes, as a result of the unavoidable attenuation of highly confined modes even if all materials are assumed perfect, with zero intrinsic bulk losses, due to confinement-induced Landau damping or nonlocality-induced radiation leakage. Our work helps clarifying the subtle and rich topological wave physics of continuous media.

\end{abstract}

\maketitle


\emph{Introduction} -- The bulk-edge correspondence is a widely used principle of topological wave physics, which allows determining the number of unidirectional edge modes from the topological properties of the bulk modes \cite{Hatsugai,Skirlo_1,Skirlo_2,Gao,Haldane_1,Haldane_2,L_Lu,S_Q_Shen,Hughes}. In topological photonic insulators with broken time-reversal symmetry (nonreciprocal), the relevant topological invariant number is the gap Chern number, i.e., the sum of the Chern numbers of all bulk modes below the bulk-mode bandgap, $ \mathcal{C}_{\mathrm{gap}}  = \sum_i \mathcal{C}_{i}  $. The bulk-edge correspondence then states that the difference in gap Chern numbers between the materials forming an interface is equal to the net number of unidirectional surface modes propagating along the interface \cite{L_Lu,S_Q_Shen,Hughes}. While this principle works well for topological photonic insulators based on periodic structures, difficulties arise in the case of continuous topological materials with no intrinsic periodicity, due to the absence of a finite Brillouin zone, which may lead to an ill-behaved response for diverging wavenumbers if spatial dispersion is not included \cite{Mario_Chern, bulk_edge_correspondence, Optica, Buddhiraju}. A proof of the bulk-edge correspondence principle in topological photonics has been recently published \cite{Mario_proof}, which does require the inclusion of spatial dispersion (nonlocality), but hinges on the assumption of an ideally dissipationless structure. As discussed in the following, however, even if all materials are assumed intrinsically lossless (no bulk damping), dissipation channels may still be present in a physical scenario. Thus, questions still remain regarding the possibility of breaking the bulk-edge correspondence in a physical system.  

\begin{figure}[h!]
	\noindent \includegraphics[width=1.0\columnwidth]{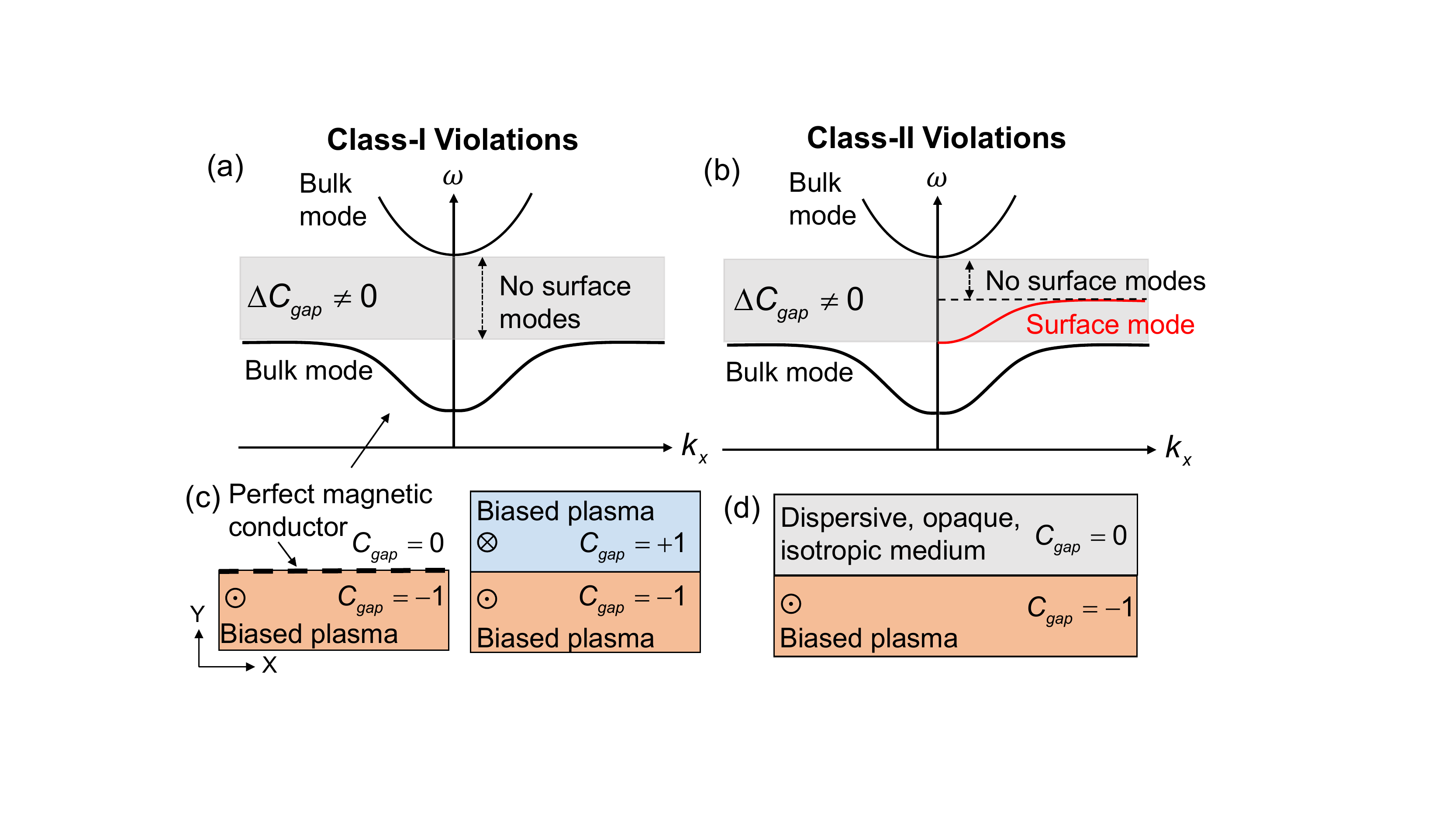}
	\caption{Apparent violations of the bulk-edge correspondence. (a,b) Illustration of the typical dispersion diagram  for the two classes of violations studied in this Letter. Solid black and red curves indicate bulk and surface modes, respectively, and gray areas highlight the bulk-mode bandgap. (c) Configurations that exhibit class-I violations of the bulk-edge correspondence. The dispersion diagram in (a) corresponds to the geometry on the left. (d) Configuration that implements a class-II violation.}
	\label{geom}
\end{figure}

Within this context, in this Letter, we present and discuss two general classes of violations of the bulk-edge correspondence in continuous topological systems, and we carefully assess the role of spatial dispersion and dissipation. These two classes are illustrated in Fig. \ref{geom}: (I) The number of unidirectional edge modes is inconsistent with the gap Chern number difference between the two materials at the interface [Fig.  \ref{geom}(a)]; (II) The dispersion curve of the unidirectional edge mode does not span the entire bulk-mode bandgap, due to the flat asymptotic dispersion of the mode at a certain frequency within the gap [Fig. \ref{geom}(b)]. Examples of these violations are easily realized using the simplest possible continuous topological photonic insulator, namely, a nonreciprocal (magnetized) plasma. Indeed, it has been known for decades that, at the interface between a magnetized plasma and a trivial photonic insulator (i.e., an opaque material, such as a conductor), unidirectional transverse-magnetic (TM) surface waves appear within the common bulk-mode bandgap \cite{Seshadri,Ishimaru}. Such surface waves are an example of unidirectional surface plasmon-polaritons (SPPs) \cite{Optica}. Recently, it has been shown that this unidirectionality is indeed rooted in the different topological properties of the trivial opaque material and the biased plasma \cite{Mario_Chern,Soljacic_1,Shen_1,Z_Yu,sink,soljacic_2,Hassani_1,Hassani_2,Zubin,Davoyan}. Specifically, considering a plasma magnetized along the $\pm z$-axis, the TM bulk modes in the plane orthogonal to the bias exhibits nontrivial topological properties, yielding a gap Chern number $ \mathcal{C}_{\mathrm{gap}} = \mp 1 $. Here, we use this model system to study physical configurations that provide apparent counterexamples to the bulk-edge correspondence. 

\emph{Class-I violations} --- To realize the simplest possible class-I violation, consider an interface between a biased plasma and a hard magnetic boundary, namely, a perfect magnetic conductor (PMC), a configuration considered for example in \cite{Ishimaru,Optica} and illustrated in Fig. \ref{geom}(c--left). Since a PMC is a topologically trivial insulator, the difference in gap Chern numbers is $ \Delta \mathcal{C}_{\mathrm{gap}} = +1 $. Thus, according to the bulk-edge correspondence principle, one topological TM surface mode should emerge across the bulk-mode bandgap. However, no TM surface mode is allowed on this interface, because the PMC boundary ``short-circuits'' the magnetic field \cite{Note1}. This absence of a surface mode within the bandgap directly violates the predictions of the bulk-edge correspondence. One might argue that, since PMCs are not natural materials, but have to be realized in the form of metasurfaces \cite{Sievenpiper}, this may be a rather artificial scenario. However, a closely related configuration that does not involve a PMC wall, but still violates the bulk-edge correspondence, is an interface between two oppositely biased plasmas [Fig. \ref{geom}(c--right)]. Indeed, as we recently showed in \cite{AWPL}, the difference in gap Chern numbers is now $ \Delta \mathcal{C}_{\mathrm{gap}} = +2 $, so one would expect two surface states, but only one appears, a mode with even-symmetric magnetic field distribution with respect to the interface. Interestingly, due to the presence of the magnetic mirror, the surface modes of the structure with the PMC wall are identical to the odd-symmetric surface modes of the configuration with oppositely biased plasmas. This suggests that the origin of the bulk-edge-correspondence violation and the nature of the missing mode are the same in both cases.



\begin{figure}[h!]
	\noindent \includegraphics[width=0.99\columnwidth]{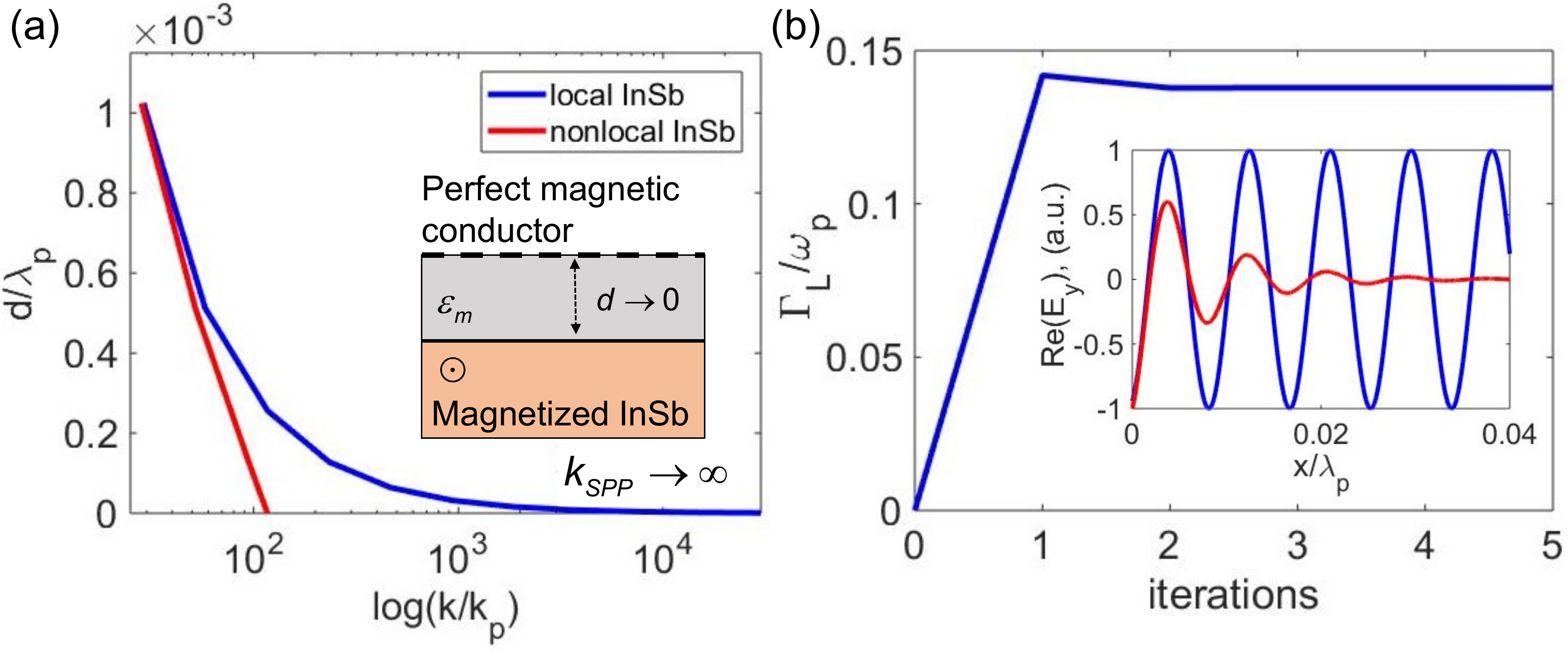}
	\caption{Class-I violations of the bulk-edge correspondence. (a) Trajectory of the surface mode wavenumber, for the configuration in the inset, as the separation $d$ between the magnetized plasma and the PMC wall is reduced to zero. Blue and red curves are the trajectories for the local and the nonlocal case ($ \beta = 1.07 \times 10^6 ~ \mathrm{m/s} $), respectively, at a frequency within the InSb bulk-mode bandgap $ \omega / \omega_p = 1.05 $. (b) Self-consistent iterative calculation of Landau damping for the configuration in (a) with $ d = 0 $, up to convergence. The inset shows the distribution (time-snapshot) of the SPP electric field normal to the interface, with zero bulk damping and zero Landau damping (blue) and with the calculated Landau damping (red).}
	\label{type_I}
\end{figure}

To clarify the nature of these violations, we consider again the configuration in Fig. \ref{geom}(c--left), but we now include a thin spacer layer, filled with an isotropic material with permittivity $ \varepsilon_m $, between the magnetized plasma and the PMC boundary, as shown in the inset of Fig. \ref{type_I}(a). The following analysis concerns TM modes, with harmonic time dependence defined by $e^{-i\omega t} $. As usually done, the magnetized plasma can be modeled by a permittivity tensor $ \boldsymbol{\varepsilon} = \varepsilon_0 \left[ \varepsilon_{11} \textbf{I}_t + \varepsilon_{33} \hat{\textbf{z}} \hat{\textbf{z}} - i\varepsilon_{12}\hat{\textbf{z}} \times \textbf{I}    \right]  $, where $ \textbf{I}_t = \textbf{I} - \hat{\textbf{z}}\hat{\textbf{z}} $. The standard dispersive models of $ \varepsilon_{11} $, $ \varepsilon_{12} $ and $ \varepsilon_{33} $ (magnetized Drude model) can be found in, e.g., \cite{Plasma}. As a realistic example of magnetized plasma, we consider the case of a magnetized $n$-doped semiconductor in the low THz regime, namely, $n$-type InSb, a material that has been studied in several recent works \cite{Palik,Boyd,Optica,Buddhiraju,Mann}. A typical sample of this material has plasma frequency $ \omega_p/2\pi = 2~ \mathrm{THz} $, electron density $ N_e = 1.1 \times 10^{22}/\mathrm{m^3} $, and bound-charge contribution to the permittivity function $ \varepsilon_{\infty} = 15.6 $. We consider a moderate DC magnetic field of 0.2 T, which is sufficient to produce a moderately large cyclotron frequency, $ \omega_c/ \omega_{p} = 0.2 $. We first consider a lossless and local scenario. Solving Maxwell's equations, with suitable boundary conditions, it is straightforward to show that the SPPs supported by the interface between the isotropic spacer and the biased plasma satisfy the following dispersion equation,
\begin{equation} \label{SPP_disp}
\frac{\varepsilon_{11} k_x + \varepsilon_{12} \alpha_p}{ \alpha_p k_x - k_0^2 \varepsilon_{12} } \cos(k_y d) - \frac{ \varepsilon_m k_y}{ k_x^2 - k_0^2 \varepsilon_{m} } \sin(k_y d) =0
\end{equation}
where $d$ is the isotropic spacer thickness, $ k_0 = \omega /c $, $ k_y = \sqrt{k_0^2\varepsilon_{m} - k_x^2} $, $ \alpha_p = \sqrt{k_x^2 - k_0^2 \varepsilon_{eff}} $ and $ \varepsilon_{eff} = \left( \varepsilon_{11}^2 - \varepsilon_{12}^2 \right)/ \varepsilon_{11} $. The solid blue curve in Fig. \ref{type_I}(a) represents the trajectory of the SPP wavenumber, calculated with \eqref{SPP_disp} at a specific frequency within the InSb bulk-mode bandgap, as the spacer thickness is reduced, which corresponds to transforming the geometry in Fig. \ref{type_I}(a) to the configuration in Fig. \ref{geom}(c--left). From this plot, one can clearly see that, as $ d \rightarrow 0 $, the SPP wavenumber tends to infinity. This analysis shows that, although the PMC-plasma configuration in Fig. \ref{geom}(c--left) does not seem to support any mode, one topological surface mode does exist in the asymptotic part of the spatial spectrum for $ k \rightarrow +\infty $. 

As mentioned above, an interface between a magnetized plasma and a PMC boundary is closely related to the case of two oppositely biased plasmas. We have recently studied this configuration in a different context \cite{AWPL}, where we showed that the wavenumber of the supported odd-symmetric mode rapidly diverges, as the spacing between the two plasmas is decreased, fully consistent with the solid blue curve trajectory in Fig. \ref{type_I}(a). This asymptotic behavior is indeed the reason for all class-I violations of the bulk-edge correspondence.

At this point, it is crucial to stress that, since we are dealing with a mode with very large wavenumber (in principle infinite), a simple material model that neglects dissipative and nonlocal effects, while mathematically consistent, is not physically accurate \cite{Raza,Agranovich}. 
We start by assessing the impact of nonlocal effects, which make the permittivity a function of the wavevector. Any plasma exhibits nonlocality mainly due to the effect of electron convection and diffusion \cite{Raza,Optica}. Specifically, here we consider a well-established hydrodynamic treatment of plasma nonlocalities; however, we stress that our considerations are general since any model of nonlocality leads to qualitatively similar predictions \cite{Ciraci,Feibelman,Esteban,Asger}. Most importantly, nonlocal effects generically leads to a high spatial-frequency cutoff for the material response \cite{Raza,Mario_Chern}.
 


Using the hydrodynamic model, the induced free-electron currents $\textbf{J}$ are governed by the following equation \cite{Raza},  
\begin{equation}\label{hydro}
\beta^2 \nabla ( \nabla \cdot \textbf{J} ) + \omega ( \omega + i\Gamma ) \textbf{J} = i\omega \left(  \omega_p^{2} \varepsilon_0 \varepsilon_{\infty} \textbf{E}  - \textbf{J} \times \omega_c \hat{z} \right)
\end{equation}
where $ \beta $ is the nonlocal parameter and $ \Gamma $ is damping rate due to absorption losses. 
By simultaneously solving Maxwell's equations and the hydrodynamic equation in the magnetized plasma, the dispersion equation of the TM bulk modes can be found as
$
k^4\beta^2 \omega^2 - k^2 \left[  \Omega_b \left(  \omega_p^2 + \beta^2k_0^2\varepsilon_{\infty}  \right) + \Omega_0     \right] + k_0^2 \varepsilon_{\infty} \Omega_0 = 0,
$
where $ \Omega_b = \omega^2 - \omega_p^2 $ and $ \Omega_0 = \Omega_b^2 - \omega^2\omega_c^2 $. If $\beta \neq 0$, the first term of this equation leads to the emergence of an additional bulk mode with respect to the local case \cite{Optica}, which implies that an additional boundary condition should be included to solve for the fields in the presence of interfaces. A physically meaningful choice is $ \textbf{J}\cdot \hat{y} = 0 $ \cite{Raza}, which forces the normal component of the induced current
density to vanish at the boundary. Following this approach, we can assess the impact of nonlocality on the surface-wave dispersion for the configuration in Fig. \ref{geom}(c--left) as $ d \rightarrow 0 $. The solid red line in Fig. \ref{type_I}(a) shows the modified trajectory of the SPP wavenumber in the presence of nonlocality. As clearly seen, while nonlocalities have little effect for large separations and small wavenumbers, they become very important at larger values of $k$: in contrast to the local case, the SPP wavenumber does not diverge, but it stops at a finite value, $ k \approx 116.6 k_p $, for $ d \rightarrow 0 $. This result shows that an SPP with finite wavenumber is indeed supported by the PMC -- nonlocal plasma interface. Hence, the bulk-edge correspondence is formally restored if nonlocal effects are properly included in the material model. 

Interestingly, the presence of this unidirectional surface mode at a PMC -- biased plasma interface (or between oppositely-biased plasmas) has so far gone unnoticed (e.g., \cite{Ishimaru,Optica}). This is, however, not surprising. Indeed, even though the mode wavenumber does not diverge in the nonlocal case, it is still very large for any realistic values of nonlocality (InSb has one of the largest available values of $\beta$ due to its small effective mass \cite{Wub}). This implies that, first, the SPP mode has very large wave impedance (proportional to $k$ for a TM mode); therefore, it is very difficult to excite due to the large impedance mismatch with any source or feeding mechanism. Most importantly, this mode is so highly confined to the interface that it becomes extremely sensitive to absorption. Thus, this highly oscillating topological mode is quickly damped in the presence of a loss channel of any type, including electron-electron interactions, phonon and defect scattering, and interface roughness (or even just the effective loss due to numerical error and mesh granularity in numerical simulations). Even in the absence of any intrinsic scattering loss in the considered materials, it is physically impossible for a surface plasmon-polariton to experience zero dissipation. Indeed, no matter the intrinsic quality of the materials and interfaces, the smallest possible loss is determined by surface-collision-induced Landau damping, i.e., by the direct excitation of electron-hole pairs in the plasma by the highly confined electric field on the interface. Following \cite{Khurgin_1,Khurgin_2}, we determine the loss rate due to Landau damping for the geometry in Fig. \ref{geom}(c--left) as
\begin{equation}\label{Landau}
\Gamma_L = \frac{3\pi \omega}{ 2 } \frac{   \int_{1}^{\infty} q^{-3} | F_y (q) |^2 dq  }{  \int_{0}^{\infty} \left( |F_x (q) |^2 + |F_y (q) |^2 \right) dq   }, 
\end{equation}
where $ F_y (q) $ and $ F_x (q) $ are the Fourier transforms of the electric-field components,  normal and parallel to the interface, and $ q = k/ (\omega / v_F) $ is a normalized wavenumber, where $ v_F $ is the Fermi velocity, proportional to the nonlocal parameter $ \beta^2 = v_F^2 \left(\frac{3}{5} \omega + \frac{1}{3}i\Gamma\right)/ \left( \omega + i\Gamma \right) $ \cite{Halevi,Goodnick}. To investigate the effect of Landau damping, we consider an ideal sample of InSb with zero intrinsic bulk loss, $ \Gamma = 0 $, and solve for $ \Gamma_L $ and the resulting SPP fields. The correct solution needs to be found in a self-consistent manner, i.e., iteratively \cite{Note 2}. Figure \ref{type_I}(b) shows the iterative solution steps for $ \Gamma_L $ up to convergence. It can be seen that, even in the absence of intrinsic bulk loss, a significant level of Landau damping is present, $ \Gamma_L \approx 0.13 \omega_p$, which represents \emph{the smallest possible physical level of loss for this configuration}. Surface-induced Landau damping strongly affects the highly confined topological surface mode. The inset of Figure \ref{type_I}(b) compares the SPP field distributions in lossless case (blue curve) and including the calculated Landau damping (red curve). The difference is striking: the surface mode dies out very quickly in the presence of Landau damping, over a distance of less than $0.03\lambda_p$ ($ \lambda_p  $ is the free-space wavelength at $\omega_p$). 

We would like to summarize here the main message of this section: While the inclusion of nonlocal effects formally restores the bulk-edge correspondence in the considered lossless configurations, the correspondence principle is physically violated since, due to confinement-induced damping, the surface mode is attenuated almost immediately even if the considered materials are assumed perfect, with zero intrinsic bulk losses.

\emph{Class-II violations} --- To realize a class-II violation of the bulk-edge correspondence, as illustrated in Fig. \ref{geom}(b), we consider an interface between a magnetized plasma and a trivial opaque medium, for example an electric conductor or an unbiased plasma, as shown in Fig. \ref{geom}(d). A large body of work \cite{Seshadri,Ishimaru,Mario_Chern,Mario_proof,Haldane_1,Haldane_2,Shen_1,Z_Yu,sink,Davoyan,bulk_edge_correspondence,AWPL,Optica,Buddhiraju,Boyd} has shown that the bulk-edge correspondence appears to correctly predict the emergence of one topological surface state within the common bulk-mode bandgap of the two materials, consistent with $ \Delta \mathcal{C}_{\mathrm{gap}} = +1 $. If the opaque material is frequency dispersive, following, for instance, a classical Drude dispersion with plasma frequency $\omega_{p}^m$, the interface supports a surface-plasmon resonance at a frequency $\omega_{SP}<\omega_{p}^m$ at which $\varepsilon_m = -(\varepsilon_{11} \pm \varepsilon_{12})$. At this frequency, the surface-mode band flattens out and tends to infinite wavenumber, as recognized in \cite{Optica}. Surface-wave propagation is not allowed at frequencies $\omega_{SP}<\omega<\omega_{p}^m$.
Thus, if we tune $\omega_{p}^m$ such that $\omega_{SP}$ falls within the bulk-mode bandgap of the magnetized plasma, as illustrated in Fig. \ref{geom}(b), we obtain a common bandgap for the two materials whose lower-frequency portion supports exactly one topological surface mode, but no surface mode can propagate in the higher-frequency portion of the bandgap. This scenario realizes another apparent violation of the bulk-edge correspondence. 

\begin{figure}[h!]
	\noindent \includegraphics[width=0.99\columnwidth]{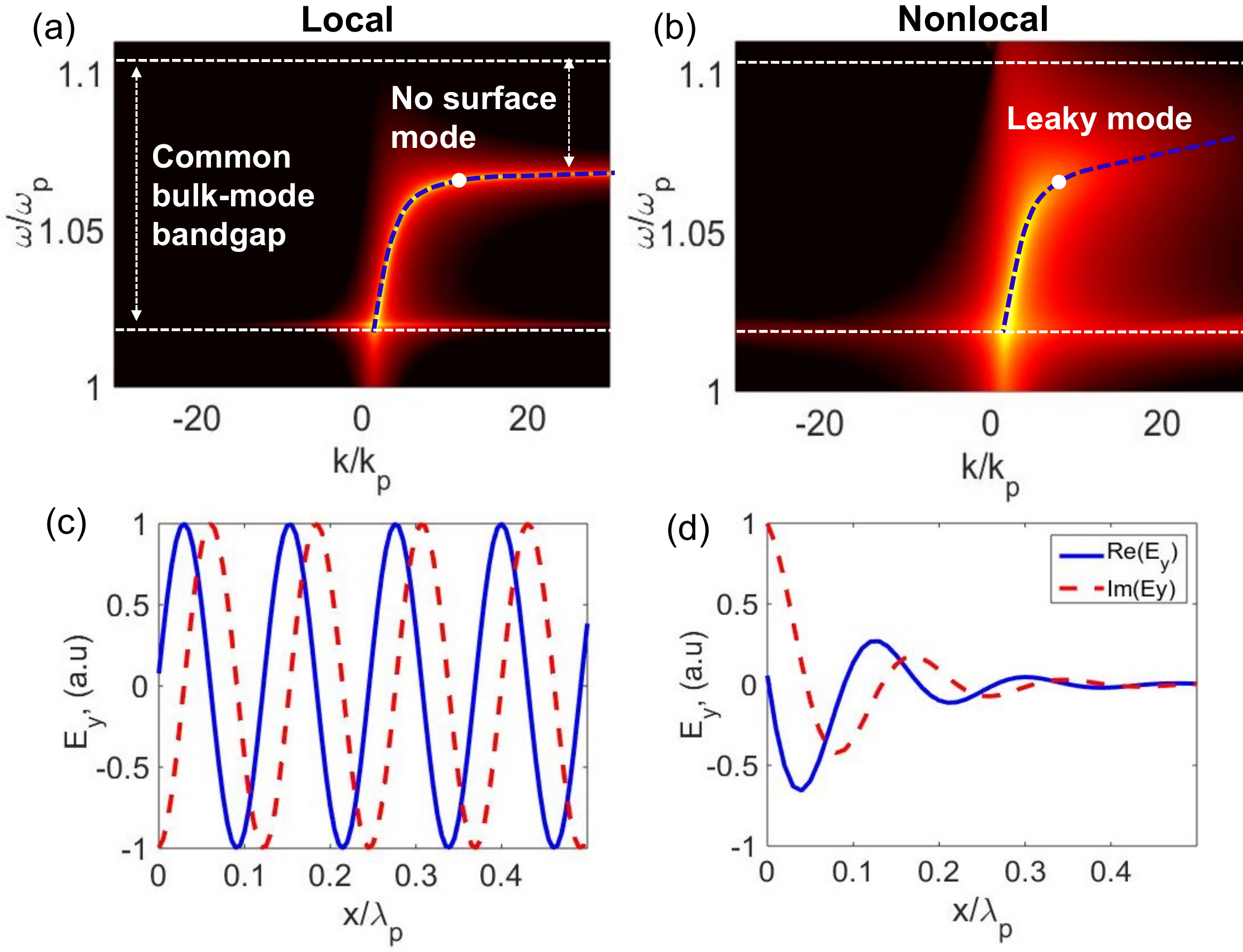}
	\caption{Class-II violations of the bulk-edge correspondence. (a,b) Dispersion diagrams (dashed blue lines) of the surface modes supported by the configuration in Fig. \ref{geom}(d). The dispersion diagrams are also plotted as density plots of the inverse determinant of the boundary-condition matrix. The bright bands correspond to the SPP poles, with the band becoming blurred if the pole moves to a complex value of wavenumber. (a) Local and lossless case, with $ \beta = \beta^m = 0 $, $ \omega_{p}^m = 1.2 \omega_p $ and $ \varepsilon_{\infty}^m = \varepsilon_{\infty} $. (b) Nonlocal and lossless case with $ \beta = \beta^m = 1.07\times 10^6 $ m/s. The dashed white lines indicate the edges of bulk-mode bandgap. (c,d) Spatial distributions (time-snapshots) of the SPP electric field normal to the interface, for the local (c) and nonlocal (d) cases, calculated at the frequency indicated by the white dot in panels (a,b).}
	\label{type_II}
\end{figure}

To implement a realistic example of class-II violation and assess the impact of nonlocal effects, we consider an interface between magnetized $n$-type InSb (same parameter as above) and a dispersive isotropic metal with parameters given in the caption of Fig. \ref{type_II}. We first consider a local scenario, and we choose the metal plasma frequency close to the upper edge of the InSb bulk-mode bandgap, for instance $ \omega_{p}^m = 1.2 \omega_p $. Fig. \ref{type_II}(a) shows the dispersion curve of the unidirectional surface mode supported by this configuration, revealing its flat asymptotic dispersion within the bandgap. As mentioned above, this behavior violates the bulk-edge correspondence, which would predict the presence of one unidirectional surface mode spanning the entire bandgap. As for class-I violations, this form of violation is due to the asymptotic behavior of the surface modes for large wavenumbers, which suggests that nonlocality may again restore the correspondence. To verify this, we include hydrodynamic nonlocalities both in the metal and the magnetized plasma, following the approach discussed in \cite{Optica,Boardman}.

If nonlocal effects are properly included, the nonphysical flat asymptotes in the modal dispersion disappear, as discussed in \cite{Optica,Buddhiraju} for different scenarios. Indeed, as shown in Fig. \ref{type_II}(b), in the nonlocal case under consideration, the dispersion curve of the unidirectional SPP monotonically grows with frequency, thus spanning the entire bulk-mode bandgap. This implies that, for each frequency within the gap, there exists exactly one unidirectional surface mode, which is now consistent with the bulk-edge correspondence. Most importantly, however, we also note that the dispersion band in Fig. \ref{type_II}(b) becomes significantly ``blurred'' at higher frequencies, which is a direct indication of the migration of the dispersion-equation root (pole of the system's Green function) into the complex wavenumber plane. Since we assumed that all the materials are lossless, a complex modal wavenumber indicates \emph{radiation loss}: the bound surface mode has become a ``leaky mode'' that gradually loses energy as radiation \cite{Francesco_IEE_Proc}. This form of nonlocality-induced radiation leakage was originally predicted in \cite{Boardman} and recently discussed in \cite{Optica}, where it was shown that the additional nonlocality-induced bulk mode of the magnetized plasma provides a continuum of radiation modes, propagating at different angles, which the surface mode can couple to, thus leaking energy into the bulk. Figs. \ref{type_II}(c,d) compare the field distribution of the SPP mode, in the local and nonlocal cases, at a frequency near the middle of the bandgap [white dot in Fig. \ref{type_II}(a,b)]: in the nonlocal case, the unidirectional mode is quickly attenuated over a distance of less than $0.5 \lambda_p$ due to radiation leakage. Even higher attenuation is obtained at higher frequencies. We also note that, because of such a strong attenuation, the degree of confinement of the SPP mode is low (especially compared to the SPPs in Fig. \ref{type_I}); therefore, the effect of Landau damping in this configuration is negligible compared to radiation damping. However, the main finding of this analysis is essentially the same as in the previous section: While the inclusion of nonlocal effects formally restores the bulk-edge correspondence, the correspondence principle is violated for all practical purposes because 
the surface mode is attenuated almost immediately even if the considered materials are assumed intrinsically lossless.

\emph{Discussion and conclusion} -- We would like to note that both classes of apparent violations discussed here arise due to the asymptotic behavior of the photonic bands for diverging wavenumbers. This is possible only for continuous media, with an infinite Brillouin zone. Conversely, topological photonic insulators based on periodic structures, e.g., photonic crystals \cite{L_Lu,Khanikaev}, are not expected to exhibit the type of violations of the bulk-edge correspondence discussed here. Even for continuous media, we have shown that the correspondence principle is formally restored if nonlocality is included in the material model, consistent with the correct definition of Chern number in uniform media, which requires a high-spatial frequency cutoff as demonstrated in \cite{Mario_Chern,bulk_edge_correspondence}. Most importantly, however, we have shown that, even if the considered materials are assumed perfect, with zero intrinsic bulk losses, the ``recovered'' surface mode that would satisfy the bulk-edge correspondence is very strongly attenuated due to either confinement-induced Landau damping, or nonlocality-induced radiation leakage. We, therefore, conclude that, in the considered cases, the bulk-edge correspondence principle is violated for all practical purposes.

We also note that our findings help clarify the behavior of some extreme nonreciprocal configurations, such as terminated one-way waveguides. In a junction between two structures, one supporting one or more unidirectional modes, and one that does not support any mode (as in the case of an interface between a magnetized plasma and a metal terminated by an orthogonal PMC wall, forming the T-like junction studied in several recent papers \cite{Chettiar,shen_2,Marvasti,Optica,Boyd,sink,Buddhiraju}), the energy incident on the junction will not be able to escape. We argue, however, that this situation does not pose any physical problem (the energy does not build up indefinitely at the junction/termination), even assuming media with vanishing losses, since the incident wave will be dissipated into a so-called ``wedge mode'' while the wavelength shrinks to zero, as discussed in \cite{Optica,Mann}, and as a result of Landau damping or radiation leakage, as discussed here.

To conclude, our work identifies, for the first time, extreme configurations in which the bulk-edge correspondence principle is physically violated. Our findings are expected to help making more accurate predictions in topological systems involving uniform media.




\begin{thebibliography}{99}
\bibitem{Note 2} At each iteration after the first one, we calculate the fields with the loss rate obtained with Eq. \eqref{Landau} at the previous step, which modifies the field intensity, the degree of mode confinement, and, therefore, the value of $ \Gamma_L $, as further explained in \cite{Khurgin_1,Khurgin_2}.
\end{thebibliography}
\end{document}